\documentclass[aps,twocolumn,showpacs,superscriptaddress,pre,A4]{revtex4}
\usepackage{amsmath}
\usepackage{amssymb}
\usepackage{mathtext}

\usepackage{graphicx}
\usepackage{mathtext}

\begin{document}

\renewcommand{\Re}{\mathrm{Re}}
\renewcommand{\Im}{\mathrm{Im}}

\title[Interfacial waves in two-layer fluid system subject to
vibrations]{Running interfacial waves in two-layer fluid system
subject to longitudinal vibrations}
\author{D.\ S.\ Goldobin}
\affiliation{Institute of Continuous Media Mechanics, UB RAS,
         1 Academik Korolev str., Perm 614013, Russia}
\affiliation{Department of Theoretical Physics,
         Perm State University, 15 Bukireva str., Perm 614990, Russia}
\affiliation{Department of Mathematics, University of Leicester,
         University Road, Leicester LE1 7RH, UK}
\author{A.\ V.\ Pimenova}
\affiliation{Institute of Continuous Media Mechanics, UB RAS,
         1 Academik Korolev str., Perm 614013, Russia}
\author{K.\ V.\ Kovalevskaya}
\affiliation{Institute of Continuous Media Mechanics, UB RAS,
         1 Academik Korolev str., Perm 614013, Russia}
\author{D.\ V.\ Lyubimov}
\affiliation{Department of Theoretical Physics,
         Perm State University, 15 Bukireva str., Perm 614990, Russia}
\author{T.\ P.\ Lyubimova}
\affiliation{Institute of Continuous Media Mechanics, UB RAS,
         1 Academik Korolev str., Perm 614013, Russia}
\affiliation{Department of Theoretical Physics,
         Perm State University, 15 Bukireva str., Perm 614990, Russia}

\begin{abstract}
We study the waves at the interface between two thin horizontal
layers of immiscible fluids subject to high-frequency horizontal
vibrations. Previously, the variational principle for energy
functional, which can be adopted for treatment of quasi-stationary
states of free interface in fluid dynamical systems subject to
vibrations, revealed existence of standing periodic waves and
solitons in this system. However, this approach does not provide
regular means for dealing with evolutionary problems: neither
stability problems nor ones associated with propagating waves. In
this work, we rigorously derive the evolution equations for long
waves in the system, which turn out to be identical to the `plus'
(or `good') Boussinesq equation. With these equations one can find
all time-independent-profile solitary waves (standing solitons are
a specific case of these propagating waves), which exist below the
linear instability threshold; the standing and slow solitons are
always unstable while fast solitons are stable. Depending on
initial perturbations, unstable solitons either grow in an
explosive manner, which means layer rupture in a finite time, or
falls apart into stable solitons. The results are derived within
the long-wave approximation as the linear stability analysis for
the flat-interface state [D.V.\ Lyubimov, A.A.\ Cherepanov,
Fluid Dynamics {\bf 21}, 849--854 (1987)] reveals the instabilities
of thin layers to be long-wavelength.


\end{abstract}

\maketitle

\section{Introduction}
\label{sec_intro}
\begin{figure}[t]
\center{
 \includegraphics[width=0.47\textwidth]%
 {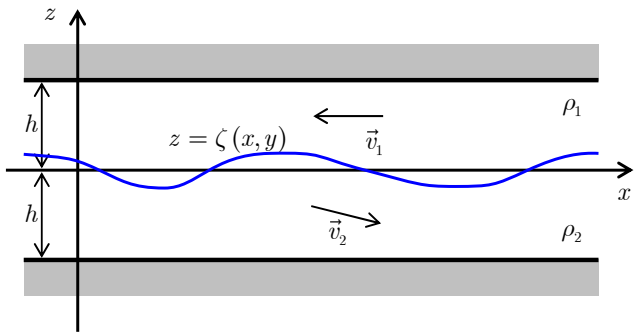}}

  \caption{
Sketch of a two-layer fluid system subject to longitudinal
vibrations and the coordinate frame.
 }
  \label{fig1}
\end{figure}

In~\cite{Wolf-1961,Wolf-1970} Wolf reported experimental
observations of the occurrence of steady wave patterns on the
interface between immiscible fluids subject to horizontal
vibrations. The build-up of the theoretical basis for these
experimental findings was initiated with the linear instability
analysis of the flat state of the
interface~\cite{Lyubimov-Cherepanov-1987,Khenner-Lyubimov-Shotz-1998,Khenner-etal-1999}
(see Fig.\,\ref{fig1} for the sketch of the system considered in these
works). Specifically, it was found that in thin layers the
instability is a long-wavelength
one~\cite{Lyubimov-Cherepanov-1987}.
In~\cite{Khenner-Lyubimov-Shotz-1998,Khenner-etal-1999}, the
linear stability was determined for the case of arbitrary
frequency of vibrations.

In spite of the substantial advance in theoretical studies, the
problem proved to require subtle approaches; a comprehensive
straightforward weakly-nonlinear analysis of the system subject to
high-frequency vibrations still remains lacking in the literature
(as well as the long-wavelength one). The approach employed
in~\cite{Lyubimov-Cherepanov-1987} can be (and was) used for
analysis of time-independent quasi-steady patterns (including
non-linear ones) only, but not the evolution of these patterns
over time. This ``restricted'' analysis of the system revealed
that quasi-steady patterns can occur both via sub- and
supercritical pitchfork bifurcations, depending on the system
parameters. Later on, specifically for thin layers, which will be
the focus of our work, the excitation of patterns was shown to be
always subcritical~\cite{Zamaraev-Lyubimov-Cherepanov-1989}
(paper~\cite{Zamaraev-Lyubimov-Cherepanov-1989} is published only
in Russian, although the result can be derived
from~\cite{Lyubimov-Cherepanov-1987} as well). Within the approach
of~\cite{Lyubimov-Cherepanov-1987,Zamaraev-Lyubimov-Cherepanov-1989}
neither time-dependent patterns nor the stability of
time-independent patterns can be analyzed. Specifically for the
case of subcritical excitation, time-independent patterns may
belong to the stability boundary between the attraction basins of
the flat-interface state and the finite-amplitude pattern state in
the phase space.~\footnote{In the idealistic dissipation-free
case, when there are no attracting states, the basins of
attraction are replaced with the basins of dynamics around
corresponding states; i.e., of running perturbations around the
flat-interface state and of evolving finite-amplitude patterns.}

In this work we accomplish the task of derivation of the governing
equations for dynamics of patterns on the interface of two-layer
fluid system within the approximation of inviscid fluids. In
Wolf's experiments~\cite{Wolf-1961,Wolf-1970}, the viscous
boundary layer in the most viscous liquid was an order of
magnitude thinner than the liquid layer, meaning the approximation
of inviscid liquid is relevant. The layer is assumed to be thin
enough for the evolving patterns to be
long-wavelength~\cite{Lyubimov-Cherepanov-1987}. With the governing
equations we analyze the dynamics of the system below the linear
instability threshold, where the system turns out to be identical
to the `plus' Boussinesq equation. The system admits soliton
solutions, these solutions are parameterized with single
parameter, soliton speed. The maximal speed of solitons equals the
minimal group velocity of linear waves in the system; the soliton
waves move always slower than the packages of linear waves.
Stability analysis reveals that the standing and slow solitons are
unstable while fast solitons are stable. The system, as the `plus'
Boussinesq equation, is known to be fully integrable.

Recently, the problem of stability of a liquid film on a
horizontal substrate subject to tangential vibrations was
addressed in the
literature~\cite{Shklyaev-Alabuzhev-Khenner-2009}. The stability
analysis for space-periodic patterns and solitary waves for the
latter system was reported in~\cite{Benilov-Chugunova-2010}. The
similarity of this problem with the problem we consider and
expected similarity of results are illusive. Firstly, for the
problem of~\cite{Shklyaev-Alabuzhev-Khenner-2009} the liquid film
is involved into oscillating motion only due to viscosity, an
inviscid liquid will be motionless over the tangentially vibrating
substrate, while in the system we consider the inviscid fluid
layers will oscillate due to motion of the lateral boundaries of
the container and fluid
incompressibility~\cite{Lyubimov-Cherepanov-1987,Khenner-Lyubimov-Shotz-1998,Khenner-etal-1999}.
Secondly, the single-film case corresponds to the case of zero
density of the upper layer in a two-layer system; in the system we
consider this is a very specific case. These dissimilarities have
their reflection in the resulting mathematical models; the governing
equations for long-wavelength patterns derived
in~\cite{Shklyaev-Alabuzhev-Khenner-2009} are of the 1st order with
respect to time and the 4th order with respect to the space
coordinate and describes purely dissipative patterns in the viscous
fluid, while the equation we will report is of 2nd order in time,
4th order in space and describes non-dissipative dynamics.


The paper is organized as follows. In Sec.\,\ref{sec_statement} we
provide a physical description and mathematical model for the system
under consideration. In Sec.\,\ref{sec_deriv} the governing equations
for long-wavelength patterns are derived and discussed. In
Sec.\,\ref{sec_solitons} soliton solutions are presented and their
stability properties are analyzed. Conclusions are drawn in
Sec.\,\ref{sec_concl}.

\section{Problem statement and governing equations}
\label{sec_statement}
We consider a system of two horizontal layers of immiscible
inviscid fluid, confined between two impermeable horizontal
boundaries (see Fig.\,\ref{fig1}). The system is subject to
high-frequency longitudinal vibrations of linear polarization; the
velocity of vibrational motion of the system is $be^{i\omega
t}+c.c.$ (here ``$c.c.$'' stands for complex
conjugate). For simplicity, we consider the case of equal
thickness, say $h$, of two layers, which is not expected to change
the qualitative picture of the system behavior~\footnote{For
problems considered
in~\cite{Lyubimov-Cherepanov-1987,Khenner-Lyubimov-Shotz-1998,Khenner-etal-1999,Zamaraev-Lyubimov-Cherepanov-1989},
moderate discrepancies between two thicknesses resulted only in
quantitative corrections.} but makes calculations simpler. The
density of upper liquid $\rho_1$ is smaller than the density of
the lower one $\rho_2$. We choose the horizontal coordinate $x$
along the direction of vibrations, the $z$-axis is vertical with
origin at the unperturbed interface between layers.

In this system, at the limit of infinitely extensive layers, the
state with flat interface $z=\zeta(x,y)=0$ is always possible. In
real layers of finite extent, the oscillating lateral boundaries
enforce liquid waves perturbing the interface; however, at a
distance from these boundaries the interface will be nearly flat
as well. For inviscid fluids, this state (the ground state)
features spatially homogeneous pulsating velocity fields
$\vec{v}_{j0}$ in both layers;
\begin{equation}
\begin{array}{c}
\displaystyle
\vec{v}_{j0}=a_j(t)\vec{e}_x,\qquad
 a_j(t)=A_je^{i\omega t}+c.c.,\\[10pt]
\displaystyle
 A_1=\frac{\rho_2 b}{\rho_1+\rho_2},\qquad
 A_2=\frac{\rho_1 b}{\rho_1+\rho_2},
\end{array}
\label{eq01}
\end{equation}
where $j=1,2$ and $\vec{e}_x$ is the unit vector of the $x$-axis.
All equations and parameters in this subsection are dimensional.
The time instant $t=0$ is chosen so that $b$ and $A_j$ are real.
The result (\ref{eq01}) follows from the condition of zero
pressure jump across the uninflected interface and the condition
of the total fluid flux through the vertical cross-section being
equal $\int_{-h}^{+h}v^{(x)}dz=2h(be^{i\omega t}+c.c.)$ (which is
due to the system motion with velocity $be^{i\omega t}+c.c.$).

Considering the flow of inviscid fluid, it is convenient to introduce
the potential $\phi_j$ of the velocity field;
\begin{equation}
\vec{v}_j=-\nabla\phi_j\,.
\label{eq02}
\end{equation}
The mass conservation law for incompressible fluid,
$\nabla\cdot\vec{v}_j=0$, yields the Laplace equation for the
potential, $\Delta\phi_j=0$. The kinematic conditions on the top
and bottom boundaries
\begin{equation}
\phi_{1z}(z=h)=\phi_{2z}(z=-h)=0
\label{eq03}
\end{equation}
and on the interface $z=\zeta(x,y)$
\begin{eqnarray}
\dot{\zeta}&=&-\phi_{1z}+\nabla\phi_1\cdot\nabla\zeta\,,
\label{eq04}
\\[5pt]
\dot{\zeta}&=&-\phi_{2z}+\nabla\phi_2\cdot\nabla\zeta
\label{eq05}
\end{eqnarray}
are also to be taken into account. (In what follows, the upper
dot stands for the time-derivative and letter in subscript denotes
partial derivative with respect to the corresponding coordinate.)
Equations (\ref{eq04}) and (\ref{eq05}) can be derived from the
condition that the points of zero value of the distance function
$F=z-\zeta(x,y)$, which correspond to the position of the
interface, move with fluid, i.e., the Lagrangian derivative
(material derivative)
 $dF/dt=\partial F/\partial t+\vec{v}\cdot\nabla{F}$
is zero on the interface:
$-\dot\zeta+v^{(z)}-\vec{v}\cdot\nabla\zeta=0$, and this holds for
both fluids.

After substitution of the potential flow, the Euler equation takes
the following form:
\[
\nabla\left(-\dot{\phi}_j+\frac{1}{2}\left(\nabla\phi_j\right)^2\right)
 =\nabla\left(-\frac{1}{\rho_j}p_j-gz\right),
\]
where $g$ is the gravity. The latter equation provides the
expression for the pressure field in the volume of two fluids for
a given flow field;
\begin{equation}
p_j=p_{j0}+\rho_j\left(\dot{\phi}_j
 -\frac{1}{2}\left(\nabla\phi_j\right)^2-gz\right).
\label{eq06}
\end{equation}

Now the stress on the interface needs to be included to make the
equation system self-contained, by providing the required boundary
conditions for $\phi_j$ on the interface between the two fluids.
The pressure jump across the interface is due to the surface
tension;
\begin{equation}
z\!=\!\zeta(x,y):
\quad
p_1-p_2 =-\alpha\nabla\cdot\vec{n}\quad
 \mbox{with  }\vec{n}=\frac{\nabla F}{\left|\nabla F\right|},
\label{eq07}
\end{equation}
where $\alpha$ is the surface tension coefficient and $\vec{n}$ is
the unit vector normal to the interface.

The system we consider does not possess any internal instability
mechanisms in the absence of vibrations (unlike,
e.g.,~\cite{Thiele-Vega-Knobloch-2006,Nepomnyashchy-Simanovskii-2013}).
Vibrations discriminate one of horizontal directions and there are
no reasons to expect that close to the threshold of
vibration-induced instabilities the excited patterns will
experience spatial modulation along the $y$-direction, which is
perpendicular to the vibration polarization direction.
Furthermore, the linear stability analysis revealed the marginal
vibration-induced instability of the flat-interface state to be
long-wavelength~\cite{Lyubimov-Cherepanov-1987,Zamaraev-Lyubimov-Cherepanov-1989}.
Hence, we restrict our consideration to the case of
$(x,z)$-geometry and the long-wavelength approximation,
$\left|{\partial_x\vec{v}}\right|\ll\left|{\partial_z\vec{v}}\right|$.

\section{Governing equations for large-scale patterns}\label{sec_deriv}
\subsection{Derivation of equations}
In this section we derive the governing equation for
long-wavelength ({\it or} large-scale) patterns. We employ the
standard multiscale method with small parameters $(T^{-1}/\omega)$
and $(h/l)$, where $T$ is the characteristic time scale of the
evolution of interface patterns (to be specified below), and $l$ is
the reference horizontal length of patterns, $\partial_x\sim
l^{-1}$. The hierarchy of small parameters and the orders of
magnitude of fields will be determined in the course of
derivation.

Within the long-wavelength approximation, the solutions to the
Laplace equation for $\phi_j(x,t)$ satisfying boundary
conditions~(\ref{eq03}) in the most general form read
\begin{align}
\phi_1=-a_1(t)x+\Phi_1(x,t)-\frac{1}{2}(h-z)^2\Phi_{1xx}(x,t)
\nonumber\\[5pt]
 {}+\frac{1}{4!}(h-z)^4\Phi_{1xxxx}(x,t)-\dots\,,
\label{eqa01}
\end{align}
\begin{align}
\phi_2=-a_2(t)x+\Phi_2(x,t)-\frac{1}{2}(h+z)^2\Phi_{2xx}(x,t)
\nonumber\\[5pt]
 {}+\frac{1}{4!}(h+z)^4\Phi_{2xxxx}(x,t)-\dots\,.
\label{eqa02}
\end{align}
Here the ground state (the flat-interface state) is represented by
the terms $-a_j(t)x$; $\Phi_j(x,t)$ describe perturbation flow,
they are as yet arbitrary functions of $x$ and $t$. After
substitution of $p_j$ from expression~(\ref{eq06}) and $\phi_j$
from expressions~(\ref{eqa01})--(\ref{eqa02}), the condition of
stress balance on the interface (\ref{eq07}) reads
\begin{widetext}
\begin{eqnarray}
\displaystyle
\hspace{-40pt}
 p_{1\infty}-p_{2\infty}
 +\rho_1\Bigg[-\dot{a}_1x+\dot{\Phi}_1-\frac{(h-\zeta)^2}{2}\dot{\Phi}_{1xx}
 -\frac{1}{2}\left(-a_1+\Phi_{1x}-\frac{(h-\zeta)^2}{2}\Phi_{1xxx}\right)^2\quad
\nonumber\\[10pt]
\displaystyle
\qquad
 {}-\frac{((h-\zeta)\Phi_{1xx})^2}{2}+\dots\Bigg]
\nonumber\\[10pt]
\displaystyle
 {}-\rho_2\Bigg[-\dot{a}_2x+\dot{\Phi}_2-\frac{(h+\zeta)^2}{2}\dot{\Phi}_{2xx}
 -\frac{1}{2}\left(-a_2+\Phi_{2x}-\frac{(h+\zeta)^2}{2}\Phi_{2xxx}\right)^2\quad
\nonumber\\[10pt]
\displaystyle
 {}-\frac{((h+\zeta)\Phi_{2xx})^2}{2}+\dots\Bigg]
\nonumber\\[10pt]
\displaystyle
 {}+(\rho_2-\rho_1)g\zeta=\alpha\frac{\zeta_{xx}}{(1+\zeta_x^2)^{3/2}}\,.
\nonumber
\end{eqnarray}
\end{widetext}
Here ``\dots'' stand for terms
$\mathcal{O}_1(\dot\Phi_jh^4/l^4)+\mathcal{O}_2(\Phi_j^2h^4/l^6)+\mathcal{O}_3(a_j\Phi_jh^4/l^5)$;
here and in what follows, $\mathcal{O}_j(Z)$ stand for unspecified
contributions of the same order of smallness as their argument
$Z$, and index $j$ is used to distinguish several nonidentical
contributions to one and the same equation. We specify the order
of the neglected terms so as to facilitate tracking the
correctness of the derivations.
The difference of constants $p_{1\infty}-p_{2\infty}$ is to be
determined from the condition that in the area of vanishing
perturbations of the pulsation flow, i.e.\ $\Phi_j(x,t)=const$,
the interface remains flat, i.e.\ $\zeta(x,t)=0$. This condition
yields
$p_{1\infty}-p_{2\infty}-(\rho_1a_1^2(t)-\rho_2a_2^2(t))/2=0$. One
can choose the following units of measurements for length:
$L=\sqrt{\alpha/[(\rho_2-\rho_1)g]}$, for time: $T=L/b$, and for
the fluid densities: $\rho_\ast$---which mean replacement
\begin{equation}
\begin{array}{c}
(x,z)\to(Lx,Lz),\quad t\to Tt,\quad \zeta\to L\zeta,\\[5pt]
\Phi_j\to(L^2/T)\Phi_j,\quad \rho_i\to\rho_\ast\rho_i
\end{array}
\label{rescaling1}
\end{equation}
in equations---and rewrite the last equation in the dimensionless
form
\begin{widetext}
\begin{equation}
\begin{array}{l}
\displaystyle
 B\Bigg[\frac{\rho_1a_1^2-\rho_2a_2^2}{2}+\rho_1\dot{\Phi}_1-\rho_2\dot{\Phi}_2
 -\frac{\rho_1(h-\zeta)^2}{2}\dot{\Phi}_{1xx}+\frac{\rho_2(h+\zeta)^2}{2}\dot{\Phi}_{2xx}
 -\frac{\rho_1}{2}\left(a_1-\Phi_{1x}+\frac{1}{2}(h-\zeta)^2\Phi_{1xxx}\right)^2 \\[15pt]
\displaystyle\quad
 {}+\frac{\rho_2}{2}\left(a_2-\Phi_{2x}+\frac{1}{2}(h+\zeta)^2\Phi_{2xxx}\right)^2
 -\frac{\rho_1}{2}\left((h-\zeta)\Phi_{1xx}\right)^2
 +\frac{\rho_2}{2}\left((h+\zeta)\Phi_{2xx}\right)^2
 +\dots\Bigg]+\zeta
 =\frac{\zeta_{xx}}{(1+\zeta_x^2)^{3/2}}\,.
\end{array}
\label{eqa03}
\end{equation}
\end{widetext}
Here the dimensionless vibration parameter
\begin{equation}
 B\equiv\frac{\rho_\ast b^2}{\sqrt{\alpha(\rho_2-\rho_1)g}}=B_0+B_1
\label{eqa04}
\end{equation}
($\rho_j$ is dimensional here), where $B_0 $ is the critical value
of the vibration parameter above which the flat-interface state
becomes linearly unstable, $B_1$ is a small deviation of the
vibration parameter from the critical value. Further, kinematic
conditions~(\ref{eq04}) and (\ref{eq05}) turn into
\begin{equation}
\dot\zeta=\left(-(h-\zeta)\Phi_{1x}+\frac{1}{3!}h^3\Phi_{1xxx}
 -a_1\zeta+\dots\right)_x ,
\label{eqa05}
\end{equation}
\begin{equation}
\dot\zeta=\left((h+\zeta)\Phi_{2x}-\frac{1}{3!}h^3\Phi_{2xxx}
 -a_2\zeta+\dots\right)_x .
\label{eqa06}
\end{equation}
Here ``\dots'' stand for
$\mathcal{O}_1(\Phi_jh^2\zeta/l^3)+\mathcal{O}_2(\Phi_jh^4/l^5)$. Equations (\ref{eqa03}), (\ref{eqa05}) and (\ref{eqa06}) form a self-contained equation
system.

It is convenient to distinguish two main time-modes in various fields: the
average over vibration period part and the pulsation part;
\[
\begin{array}{l}
 \zeta=\eta(\tau,x)+\xi(\tau,x)e^{i\omega t}+c.c.+\dots\,, \\[10pt]
 \Phi_j=\varphi_j(\tau,x)+\psi_j(\tau,x)e^{i\omega t}+c.c.+\dots\,,
\end{array}
\]
where $\tau$ is a ``slow'' time related to the average over
vibration period evolution and ``\dots'' stand for higher powers
of $e^{i\omega t}$.

In order to develop an expansion in small parameter $\omega^{-1}$, we
have to adopt a certain hierarchy of smallness of parameters,
fields, etc. We adopt small deviation from the instability
threshold $B_1\sim\omega^{-1}$. Then $\eta\sim\omega^{-1}$ and
$\partial_x\sim\omega^{-1/2}$
(cf.~\cite{Lyubimov-Cherepanov-1987,Zamaraev-Lyubimov-Cherepanov-1989}).
It is as well established (e.g.,
\cite{Lyubimov-Cherepanov-1987,Zamaraev-Lyubimov-Cherepanov-1989})
that for finite wavelength perturbations (finite $k\ne0$)
$B_0(k)=B_0(0)+Ck^2+\mathcal{O}(k^4)$. Generally, the expansion of
the exponential growth rate of perturbations in the series for
$B_1$ near the instability threshold possesses a non-zero linear
part, and $B_0(k)-B_0(0)\sim k^2$; therefore,
$\partial_\tau\sim\mathcal{O}_1(B_1)+\mathcal{O}_2(k^2)\sim\omega^{-1}$.
It is more convenient to determine the order of magnitude of
$\xi$, $\varphi_j$ and $\psi_j$ in the course of development of
the expansion.

Collecting terms with
$e^{i\omega t}$ in equations~(\ref{eqa05}) and (\ref{eqa06}), one finds
\begin{eqnarray}
i\omega\xi+\xi_\tau=\Big(-(h-\eta)\psi_{1x}+\frac{1}{3!}h^3\psi_{1xxx}\qquad
\nonumber\\[5pt]
 {}+\xi\varphi_{1x}-A_1\eta+\dots\Big)_x\,,
\label{eqa07}
\end{eqnarray}
\begin{eqnarray}
i\omega\xi+\xi_\tau=\Big((h+\eta)\psi_{2x}-\frac{1}{3!}h^3\psi_{2xxx}\qquad
\nonumber\\[5pt]
 {}+\xi\varphi_{2x}-A_2\eta+\dots\Big)_x\,,
\label{eqa08}
\end{eqnarray}
where ``\dots'' stand for
$\mathcal{O}_1((\xi\varphi+\eta\psi)h^2/l^4)+\mathcal{O}_2(\psi\,h^4/l^6)$.
Terms constant with respect to $t$ sum up to
\begin{equation}
\eta_\tau=\Big(-(h-\eta)\varphi_{1x}+\xi\psi_{1x}^\ast+c.c.-A_1\xi^\ast+c.c.+\dots\Big)_x\,,
\label{eqa09}
\end{equation}
\begin{equation}
\eta_\tau=\Big((h+\eta)\varphi_{2x}+\xi\psi_{2x}^\ast+c.c.-A_2\xi^\ast+c.c.+\dots\Big)_x\,,
\label{eqa10}
\end{equation}
where the superscript ``$*$'' stands for complex conjugate and
``\dots'' stand for
$\mathcal{O}_1((\eta\varphi+\xi\psi)h^2/l^4)+\mathcal{O}_2(\varphi
h^4/l^6)$. The difference of equations~(\ref{eqa07}) and
(\ref{eqa08}) yields $\psi_j\sim\omega^{-1/2}$, and the difference
of (\ref{eqa09}) and (\ref{eqa10}) yields
$\varphi_j\sim\omega^{-1}$. For dealing with non-linear terms in
what follows, it is convenient to extract the first correction to
$\psi_j$ explicitly, i.e.\ write
$\psi_j=\psi_j^{(0)}+\psi_j^{(1)}+\dots$\,, where
$\psi_j^{(1)}\sim\omega^{-1}\psi _j^{(0)}\sim\omega^{-3/2}$.
Equation~(\ref{eqa07}) (or (\ref{eqa08})) yields in the leading
order ($\sim\omega^{-3/2}$)
\begin{equation}
\xi=\frac{i}{\omega}(h\psi_{1x}+A_1\eta)_x\sim\omega^{-\frac{5}{2}}\,.
\label{eqa11}
\end{equation}

Considering the difference of (\ref{eqa08}) and (\ref{eqa07}), one
has to keep in mind that we are interested in localized patterns
for which $\Phi_{jx}(x=\pm\infty)=0$, $\zeta(x=\pm\infty)=0$.
Hence, this difference can be integrated with respect to $x$,
taking the form
\begin{align}
h(\psi_1+\psi_2)_x-\eta(\psi_1-\psi_2)_x
 -\frac{1}{6}h^3(\psi_1+\psi_2)_{xxx}\qquad
\nonumber\\[5pt]
 {}-\xi(\varphi_1-\varphi_2)_x+(A_1-A_2)\eta+\dots=0\,,
\nonumber
\end{align}
which yields in the first two orders of smallness
\begin{equation}
h(\psi_1^{(0)}+\psi_2^{(0)})_x=-(A_1-A_2)\eta=-\frac{\rho_2-\rho_1}{\rho_2+\rho_1}\eta\,,
\label{eqa12}
\end{equation}
\begin{equation}
h(\psi_1^{(1)}+\psi_2^{(1)})_x=(\psi_1^{(0)}-\psi_2^{(0)})_x\eta
+\frac{1}{6}h^3(\psi_1^{(0)}+\psi_2^{(0)})_{xxx}\,.
\label{eqa13}
\end{equation}
The difference and the sum of equations~(\ref{eqa09}) and
(\ref{eqa10}) yield in the leading order, respectively,
\begin{equation}
\varphi_1=-\varphi_2\equiv\varphi\,,
\label{eqa14}
\end{equation}
\begin{equation}
\eta_\tau=-h\varphi_{xx}\,.
\label{eqa15}
\end{equation}

Let us now consider equation~(\ref{eqa03}). We will collect groups
of terms with respect to power of $e^{i\omega t}$ and the order of
smallness in $\omega^{-1}$.
\\
$\underline{\sim\omega^{+\frac{1}{2}}e^{i\omega t}}$:
$$
i\omega B_0(\rho_1\psi_1^{(0)}-\rho_2\psi_2^{(0)})=0\,.
$$
We introduce
\begin{equation}
\psi^{(0)}\equiv\rho_j\psi_j^{(0)}\,.
\label{eqa16}
\end{equation}
The last equation and equation~(\ref{eqa12}) yield
\begin{equation}
\psi_x^{(0)}=-\frac{1}{h}\frac{\rho_1\rho_2(\rho_2-\rho_1 )}{(\rho_2+\rho_1)^2}\eta\,.
\label{eqa17}
\end{equation}
$\underline{\sim\omega^0e^{i\omega t}}$:
$$
\mbox{No contributions.}
$$
$\underline{\sim\omega^{-\frac{1}{2}}e^{i\omega t}}$:
\[
\begin{array}{l}
\displaystyle
 i\omega B_1\underbrace{(\rho_1\psi_1^{(0)}-\rho_2\psi_2^{(0)})}_{\quad=0}
 +i\omega B_0(\rho_1\psi_1^{(1)}-\rho_2\psi_2^{(1)})
\\[20pt]
\displaystyle\qquad
 {}+B_0\underbrace{(\rho_1\psi_1^{(0)}-\rho_2\psi_2^{(0)})_\tau}_{\quad=0}
\\[20pt]
\displaystyle\qquad\qquad
 {}+i\omega B_0\frac{h^2}{2}\underbrace{(\rho_2\psi_{2xx}^{(0)}-\rho_1\psi_{1xx}^{(0)})}_{\quad=0}=0\,.
 \end{array}
\]
(We marked the combinations which are known to be zero from the
leading order of expansion.) Similarly to~(\ref{eqa16}), we
introduce
\begin{equation}
\psi^{(1)}\equiv\rho_j\psi_j^{(1)}\,.
\label{eqa18}
\end{equation}
The last equation and equation~(\ref{eqa13}) yield
\begin{eqnarray}
&&\hspace{-20pt}\displaystyle
\psi_x^{(1)}=\frac{1}{h}\frac{\rho_2-\rho_1}{\rho_2+\rho_1}\psi_x^{(0)}\eta
 +\frac{h^2}{6}\psi_{xxx}^{(0)}
 \nonumber\\[10pt]
&&\displaystyle
 =-\frac{\rho_1\rho_2(\rho_2-\rho_1)^2}{h^2(\rho_2+\rho_1)^3}\eta
 -\frac{h\rho_1\rho_2(\rho_2-\rho_1)}{6(\rho_2+\rho_1)^2}\eta_{xx}\,.
\label{eqa19}
\end{eqnarray}
$\underline{\sim\omega^{-1}(e^{i\omega t})^0}$:
$$
B_0[-\rho_2(A_2\psi_{2x}^{(0)\ast}+c.c.)+\rho_1(A_1\psi_{1x}^{(0)\ast}+c.c.)]+\eta=0\,.
$$
Substituting~(\ref{eqa16}) and (\ref{eqa17}) into the last
equation gives
$$
\left[-\frac{2B_0\rho_1\rho_2(\rho_2-\rho_1)^2}{h(\rho_2+\rho_1)^3}+1\right]\eta=0\,.
$$
Thus we obtain the solvability condition, which poses a
restriction on $B_0$; this restriction determines the linear
instability threshold
\begin{equation}
B_0=\frac{(\rho_2+\rho_1)^3h}{2\rho_1\rho_2(\rho_2-\rho_1)^2}\,.
\label{eqa20}
\end{equation}
$\underline{\sim\omega^{-2}(e^{i\omega t})^0}$: (using
(\ref{eqa14}) for $\varphi_j$)
$$
\begin{array}{l}
\displaystyle
 B_1\underbrace{[-\rho_2(A_2\psi_{2x}^{(0)\ast}+c.c.)
                +\rho_1(A_1\psi_{1x}^{(0)\ast}+c.c.)]}_{\qquad=-\eta/B_0}\\[5pt]
\displaystyle
 \qquad
 {}+B_0\bigg[(\rho_2+\rho_1)\varphi_\tau+\rho_2|\psi_{2x}^{(0)}|^2
\\[15pt]
\displaystyle
 {}-\rho_2\Big(A_2\psi_{2x}^{(1)\ast}+c.c.-A_2\frac{h^2}{2}\psi_{2xxx}^{(0)\ast}+c.c.\Big)
   -\rho_1|\psi_{1x}^{(0)}|^2
\\[15pt]
\displaystyle\quad
 {}+\rho_1\Big(A_1\psi_{1x}^{(1)\ast}+c.c.
 -A_1\frac{h^2}{2}\psi_{1xxx}^{(0)\ast}+c.c.\Big)\bigg]=\eta_{xx}\,.
 \end{array}
$$
Substituting $\psi_j^{(n)}$ from (\ref{eqa16})--(\ref{eqa19}), one
can rewrite the latter equation as
$$
\begin{array}{r}
\displaystyle
 -\frac{B_1}{B_0}\eta+B_0\Bigg[(\rho_2\!+\!\rho_1)\varphi_\tau
 -\frac{\rho_2\!-\!\rho_1}{\rho_2\rho_1}\left(\!\frac{\rho_1\rho_2(\rho_2\!-\!\rho_1)\eta}{h(\rho_2\!+\!\rho_1)^2}\!\right)^2
\\[15pt]
\displaystyle
 {}-\frac{2\rho_1\rho_2(\rho_2-\rho_1)^3}{h^2(\rho_2+\rho_1)^4}\eta^2
   -\frac{h\rho_1\rho_2(\rho_2-\rho_1)^2}{3(\rho_2+\rho_1)^3}\eta_{xx}
 \qquad\\[15pt]
\displaystyle
 {}+h^2\frac{\rho_1\rho_2(\rho_2-\rho_1)^2}{h(\rho_2+\rho_1)^3}\eta_{xx}\Bigg]=\eta_{xx}\,.
 \end{array}
$$
Together with equation~(\ref{eqa15}) the latter equation form the
final {\em system of governing equations for long-wavelength
perturbations} of the flat-interface state:
\begin{equation}
\left\{
 \begin{array}{rcl}
 \displaystyle
 B_0(\tilde\rho_2\!+\!\tilde\rho_1)\tilde\varphi_{\tilde\tau}&
 \displaystyle \!\!=\!\!&
 \displaystyle \left[1\!-\!\frac{\tilde{h}^2}{3}\right]\tilde\eta_{\tilde x\tilde x}
  +\frac{3}{2\tilde h}\frac{\tilde\rho_2\!-\!\tilde\rho_1}{\tilde\rho_2\!+\!\tilde\rho_1}\tilde\eta^2
  +\frac{B_1}{B_0}\tilde\eta\,,\\[15pt]
  \displaystyle
 \tilde\eta_{\tilde\tau}&
 \displaystyle \!\!=\!\!&
 \displaystyle -\tilde h\tilde\varphi_{\tilde x\tilde x}\,.
 \end{array}
\right.
\label{eq08}
\end{equation}
Here we explicitly mark the dimensionless variables and parameters
with the tilde sign to distinguish them from original dimensional
variables and parameters. Above in this paragraph, the tilde sign
was omitted to make calculations possibly less laborious. For
convenience we explicitly specify how to read
rescaling~(\ref{rescaling1}) with the tilde-notation: $x=L\tilde
x$, $t=(L/b)\tilde t$, $\rho_i=\rho_\ast\tilde\rho_i$, etc. The
expression for $B_0$ (\ref{eqa20}) in the original dimensional terms
reads
\begin{equation}
 B_0=\frac{\rho_\ast(\rho_2+\rho_1)^3h}{2\rho_1\rho_2(\rho_2-\rho_1)^2}
 \sqrt{\frac{(\rho_2-\rho_1)g}{\alpha}}\,.
\label{eq10}
\end{equation}

We remark that equation system~(\ref{eq08}) is valid for $B_1$ small
compared to $B_0$, otherwise one cannot stay within the
long-wavelength approximation. On rare occasions it is possible to use
long-wavelength for finite deviations from the linear instability
threshold and derive certain information on the system dynamics
(e.g., in~\cite{Goldobin-Lyubimov-2007} for Soret-driven
convection from localized sources of heat or solute in a thin
porous layer, an unavoidable appearance of patterns similar to
hydraulic jumps~\cite{Watson-1964} was predicted within the
long-wavelength approximation though for a finite deviation from
the linear instability threshold).

\subsection{On the long-wavelength character of the linear instability}
In the text above, we relied on the fact that instability is
long-wavelength for thin enough layers. Now we have appropriate
quantifiers to specify quantitatively, what``thin enough'' means.
According to~\cite{Lyubimov-Cherepanov-1987}, we require that $\tilde
h<\sqrt{3}$. Remarkably, we can see a footprint of this fact from
equation system~(\ref{eq08}) with multiplier $[1-\tilde{h}^2/3]$
ahead of $\tilde\eta_{\tilde x\tilde x}$. Indeed, the exponential
growth rate $\tilde\lambda$ of linear normal perturbations
$(\tilde\eta,\tilde\varphi)\propto\exp(\tilde\lambda\tilde
t+i\tilde k\tilde x)$ of the trivial state obeys
\begin{equation}
\tilde\lambda^2=\frac{\tilde h\,\tilde{k}^2}{B_0(\tilde\rho_2+\tilde\rho_1)}
\left(-\Bigg[1-\frac{\tilde{h}^2}{3}\Bigg]\tilde{k}^2+\frac{B_1}{B_0}\right).
\label{eq11}
\end{equation}
Below the linear instability threshold of infinitely long
wavelength perturbations, i.e.\ for $B_1<0$, there are no growing
perturbations for $\tilde{h}<\sqrt{3}$, while the perturbations
with large enough $\tilde k$ grow for $\tilde{h}>\sqrt{3}$. Of
course, this analysis of equation system~(\ref{eq08}) only
highlights the long-wavelength character of the linear
instability, since it deals with the limit of small $\tilde{k}$
and does not provide information on the linear stability for
finite $\tilde{k}$. A comprehensive proof of the long-wavelength
character of the instability for $\tilde{h}<\sqrt{3}$ comes
from~\cite{Lyubimov-Cherepanov-1987}.

In the following we will consider system behavior below the
linear instability threshold, i.e.\ for negative $B_1$. It is
convenient to make further rescaling of coordinates and variables:
\begin{equation}
\begin{array}{l}
\displaystyle
 \tilde x\to x\sqrt{\frac{B_0}{(-B_1)}\Bigg[1-\frac{\tilde{h}^2}{3}\Bigg]}\,,
\\[20pt]
\displaystyle
 \tilde t\to t\sqrt{\frac{\tilde\rho_2-\tilde\rho_1}{\tilde{h}}\frac{B_0^3}{B_1^2}
 \Bigg[1-\frac{\tilde{h}^2}{3}\Bigg]}\,,\\[20pt]
\displaystyle
 \tilde\eta\to\eta\,\tilde{h}\frac{\tilde\rho_2+\tilde\rho_1}{\tilde\rho_2-\tilde\rho_1}\frac{(-B_1)}{B_0}\,,
\\[20pt]
\displaystyle
 \tilde\varphi\to\varphi\sqrt{\frac{(\tilde\rho_2+\tilde\rho_1)^2}{(\tilde\rho_2-\tilde\rho_1)^3}
 \frac{B_1^2}{\tilde{h}B_0^3}\Bigg[1-\frac{\tilde{h}^2}{3}\Bigg]}\,.
 \end{array}
\label{eq12}
\end{equation}
We note that this implies the following rescaling of {\it initial
dimensional} coordinates and variables:
\begin{equation}
\begin{array}{l}
\displaystyle
 x\to x\,L\sqrt{\frac{B_0}{(-B_1)}\Bigg[1-\frac{h^2}{3L^2}\Bigg]}\,,\\[20pt]
\displaystyle
 t\to t\sqrt{\frac{\rho_2-\rho_1}{\rho_\ast}\frac{L^3B_0^3}{h\,b^2B_1^2}
 \Bigg[1-\frac{h^2}{3L^2}\Bigg]}\,,\\[20pt]
\displaystyle
 \eta\to\eta\,h\frac{\rho_2+\rho_1}{\rho_2-\rho_1}\frac{(-B_1)}{B_0}\,,\\[20pt]
\displaystyle
 \varphi\to\varphi\sqrt{\frac{\rho_\ast(\rho_2+\rho_1)^2}{(\rho_2-\rho_1)^3}
 \frac{L^3B_1^2}{h\,b^2B_0^3}\Bigg[1-\frac{h^2}{3L^2}\Bigg]}\,.
 \end{array}
\label{eq13}
\end{equation}
After this rescaling, equation system~(\ref{eq08}) takes
the zero-parametric form;
\begin{eqnarray}
\dot\varphi&=&\eta_{xx}+\frac{3}{2}\eta^2-\eta\,,
\label{eq14}
\\
\dot\eta&=&-\varphi_{xx}\,.
\label{eq15}
\end{eqnarray}

The derivation of the latter equation system itself is one of the
main results we report with this paper, as it allows consideration of
the evolution of quasi-steady patterns in the two-layer fluid
system under the action of the vibration field.

\subsection{The `plus' Boussinesq equation and the original Boussinesq
            equation for gravity waves in shallow water}
The equation system (\ref{eq14})--(\ref{eq15}) can be rewritten in
the form of a `plus' Boussinesq equation (plus BE);
\begin{equation}
\ddot{\eta}-\eta_{xx}+\left(\frac{3}{2}\eta^2+\eta_{xx}\right)_{xx}=0\,.
\label{eq_plBE}
\end{equation}
Meanwhile, the original Boussinesq equation B (BE~B) for gravity
waves in a shallow water layer~\cite{Boussinesq-1872} or in a
two-layer system without vibrations~\cite{Choi-Camassa-1999} reads
\begin{equation}
\ddot{\eta}-\eta_{xx}-\left(\frac{3}{2}\eta^2+\eta_{xx}\right)_{xx}=0\,.
\label{eq_BEB}
\end{equation}
Both systems are fully integrable and multi-soliton solutions are
known for them from the literature
(e.g.,~\cite{Manoranjan-etal-1984,Manoranjan-etal-1988,Bogdanov-Zakharov-2002}).
However, their dynamics is essentially different; the original
BE~B suffers from the short-wave instability, while the plus BE is
free from this instability. Solitons in the plus BE can be
unstable, decaying into pairs of stable solitons or experiencing
explosive formation of sharp peaks in finite
time~\cite{Bogdanov-Zakharov-2002,Bona-Sachs-1988,Liu-1993}. In
the sections below we will provide overview of the soliton
dynamics for equation~(\ref{eq_plBE}) in relation to the fluid
dynamical system we deal with. Prior to doing so, in this
subsection, we would like to focus more on discussion of different
kinds of the generalized Boussinesq equation and their
relationships with dynamics of systems of fluid layers.

Small-amplitude gravity waves in shallow water are governed by the
set A Boussinesq equations (equation system (25)
in~\cite{Boussinesq-1872}) which read in our terms after proper
rescaling as
\begin{equation}
\left\{
\begin{array}{rcl}
\displaystyle\dot{\eta}+\varphi_{xx}&\!\!=\!\!&
 \displaystyle-(\eta\,\varphi_x)_x+\frac{1}{6}\varphi_{xxxx}\,,
\\[10pt]
\displaystyle\dot{\varphi}+\eta&\!\!=\!\!&
 \displaystyle-\frac{1}{2}(\varphi_x)^2+\frac{1}{2}\dot\varphi_{xx}\,,
\end{array}
\right.
\end{equation}
where the terms in the right hand side of equations are small,
i.e., both nonlinearity and dispersion are small. To the leading
corrections pertaining to nonlinearity and dispersion, the latter
equation system can be recast as
\begin{equation}
\ddot{\eta}-\eta_{xx}=\left((\varphi_x)^2+\frac{1}{2}\eta^2+\eta_{xx}\right)_{xx}\,,
\end{equation}
where small terms are collected in the r.h.s.\ part of the
equation. For waves propagating in one direction
$\partial_x\approx\pm\partial_t$ and, to the leading corrections,
one can make substitution
$(\varphi_x)^2\approx(\dot\varphi)^2\approx\eta$, which yields
equation~(\ref{eq_BEB}). Thus, the Boussinesq equation for the
classical problem of waves in shallow water is not only inaccurate
far from the edge of the spectrum of soliton speed (near $c=1$)
but is also inappropriate for consideration of collisions of
counterpropagating waves (as $|\varphi_x|\ne|\dot\varphi|$ for
them). In contrast, the equations we derived for our physical
system are accurate close to the vibration-induced instability
threshold for the entire range of soliton speeds and all kinds of
soliton interactions as long as the profile remains smooth.

It is also noteworthy, that character of the original Boussinesq
equation B is inherent to the dynamics of inviscid fluid layers in
force fields and does not change without special external fields,
the action of which cannot be formally represented by any
correction to the gravity. The case of a vibration field turns out
to be one such and yields the dynamics governed by the plus BE.

\section{Long waves below the linear instability threshold}
\label{sec_solitons}
In this section we consider waves in the dynamic
system~(\ref{eq14})--(\ref{eq15}). In equations~(\ref{eq13}), one
can see how the rescaling of each coordinate and variable depends
on $(-B_1)=B_0-B$. From these dependencies it can be seen, that
for patterns in the dynamic system~(\ref{eq14})--(\ref{eq15}) the
corresponding patterns in real time--space will obey the following
scaling behavior near the linear instability threshold: spatial
extent $x_\ast\propto1/\sqrt{B_0-B}$, reference time
$t_\ast\propto1/(B_0-B)$, reference profile deviation
$\eta_\ast\propto(B_0-B)$.

\begin{figure}[t]
\center{
 \includegraphics[width=0.40\textwidth]%
 {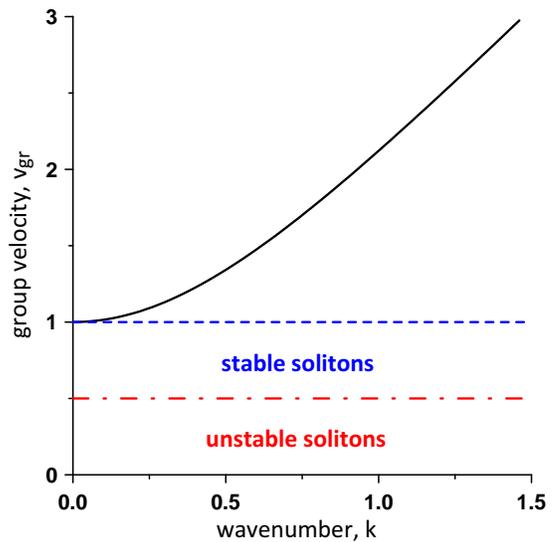}}

  \caption{
Group velocity $v_\mathrm{gr}$ of linear waves (black solid line)
vs the wavenumber $k$. The group velocity for all wave packages is
larger than the maximal soliton propagation velocity $c=1$.
(Soliton stability is discussed in the text below.)
 }
  \label{fig2}
\end{figure}

\subsection{Linear waves: dispersion equation, group velocity}
Let us first describe propagation of small perturbation, linear
waves, in the dynamic system~(\ref{eq14})--(\ref{eq15}). For normal
perturbations $(\eta,\varphi)\propto\exp(-i\Omega t+ikx)$ the
oscillation frequency reads
\begin{equation}
\Omega(k)=k\sqrt{1+k^2}\,.
\label{eq16}
\end{equation}
The corresponding phase velocity is
\begin{equation}
v_\mathrm{ph}=\Omega/k=\sqrt{1+k^2}\,,
\label{eq17}
\end{equation}
and the group velocity, which describes propagation of envelopes
of wave packages, is
\begin{equation}
v_\mathrm{gr}=\frac{\mathrm{d}\Omega}{\mathrm{d}k}=\frac{1+2k^2}{\sqrt{1+k^2}}\,.
\label{eq18}
\end{equation}
One can see that the minimal group velocity is 1 and the group
velocity $v_\mathrm{gr}$ monotonously increases as wavelength
decreases (see Fig.\,\ref{fig2}).

\begin{figure*}[t]
\center{
 \includegraphics[width=0.18\textwidth]%
 {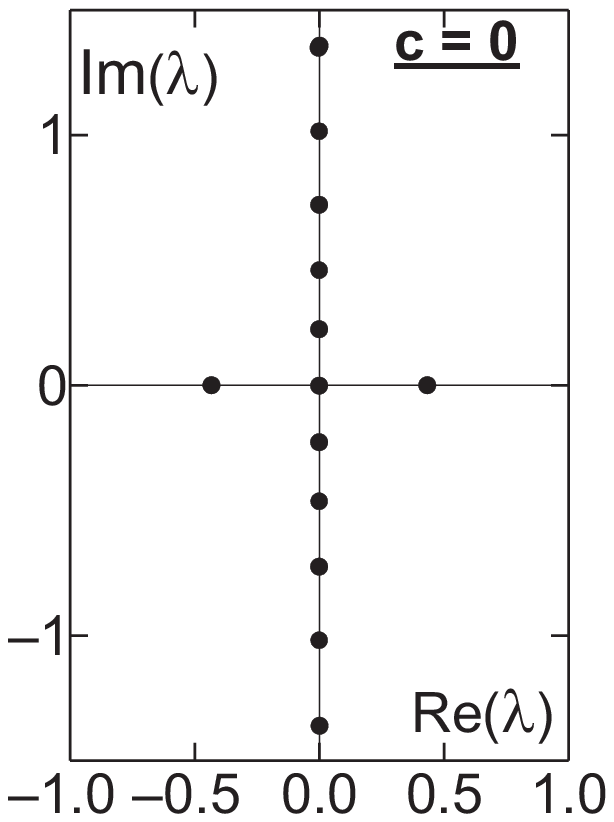}
 \qquad
 \includegraphics[width=0.18\textwidth]%
 {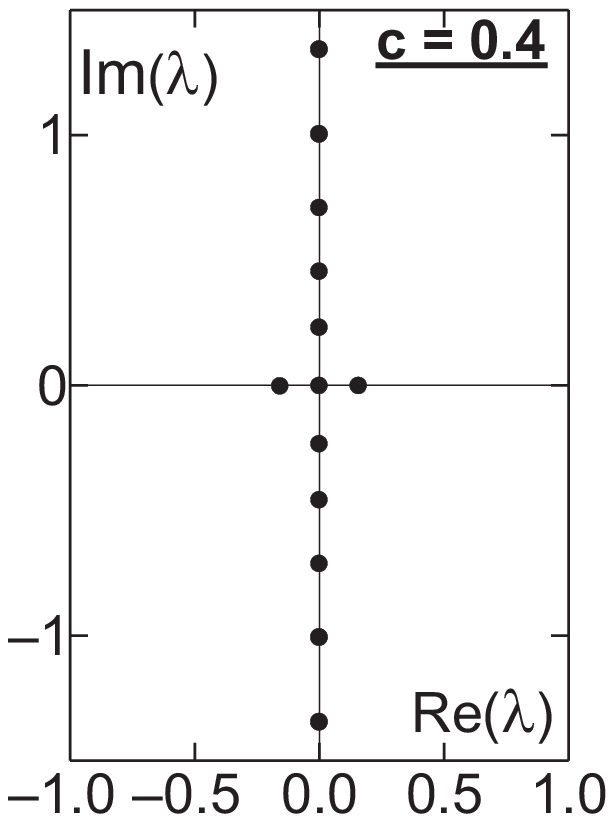}
 \qquad
 \includegraphics[width=0.18\textwidth]%
 {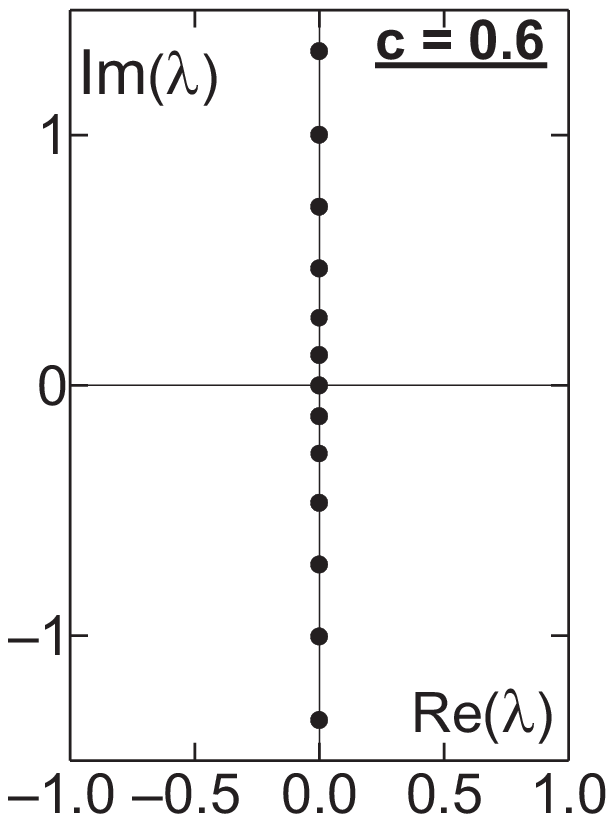}
 \qquad
 \includegraphics[width=0.18\textwidth]%
 {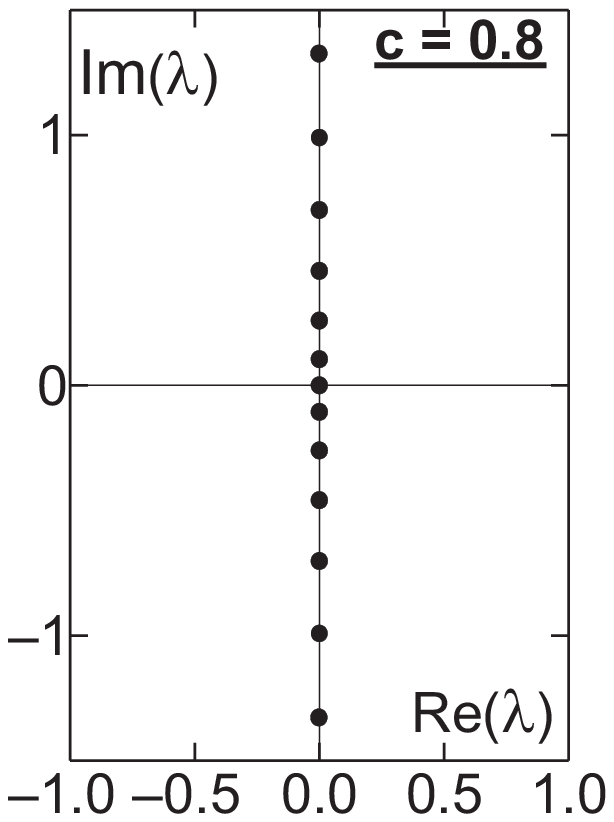}}

  \caption{Spectra of eigenvalues $\lambda$
of the problem~(\ref{eq22})--(\ref{eq24}) of linear stability of
solitons with speed $c$ specified in plots. The real part of
$\lambda$ is the exponential growth rate of perturbations; for
$c=0$ and $c=0.4$, one can see existence of one mode with positive
$\Re(\lambda)$, i.e.\ instability mode, while for $c=0.6$ and
$c=0.8$, all the perturbations oscillate without growth. In
Fig.\,\ref{fig4}(b), the instability mode and the Goldstone mode of
neutral stability (invariance to the shifts of soliton in space)
are plotted for standing solitons ($c=0$).
 }
  \label{fig3}
\end{figure*}

\subsection{Solitons}
The dynamic system~(\ref{eq14})--(\ref{eq15}) admits
time-independent-profile solutions, solitons
$\eta(x,t)=\eta(x-ct)$, where $c$ is the soliton velocity. With
identical equality $\partial_t\eta(x-ct)=-c\partial_x\eta(x-ct)$,
for localized patterns, which vanish at $x\to\pm\infty$, equation
(\ref{eq15}) can be once integrated and yields
$\varphi^\prime=c\eta$ (here the prime denotes the differentiation
with respect to argument).
Eq.\,(\ref{eq14}) takes the form
\begin{equation}
0=\eta^{\prime\prime}+\frac32\eta^2-(1-c^2)\eta\,.
\label{eq19}
\end{equation}
The latter equation admits the soliton solution
\begin{equation}
\eta_0(x,t)=\frac{1-c^2}{\displaystyle\cosh^2\frac{\sqrt{1-c^2}(x\pm ct)}{2}}\,,
\label{eq20}
\end{equation}
the propagation direction ($+c$ or $-c$) is determined by the
flow, $\varphi^\prime=\pm c\eta$. The family of soliton solutions
turns out to be one-parametric, parameterized by the speed $c$
only. The speed $c$ varies within the range $[0,1]$; standing soliton
($c=0$) is the sharpest and the tallest one and for the fastest
solitons, $c\to1$, the spatial extent tends to infinity, while the
height tends to $0$.

Considering in the same way a non-rescaled equation
system~(\ref{eq08}), one can see, that for a given physical system
with vibration parameter $B$ as a control parameter, the shape of
a soliton solution is controlled by combination
\begin{equation}
[(-B_1)/B_0-\tilde{c}^2\tilde{h}^{-1}B_0(\tilde\rho_2+\tilde\rho_1)]\,.
\label{eq21}
\end{equation}
This means that one and the same interface inflection soliton can
exist for different values of $B$, though, since the
shape-controlling parameter~(\ref{eq21}) should be the same, the
non-rescaled soliton run speed $\tilde{c}$ grows as the departure
from the threshold $(-B_1)$ increases.

Since $v_\mathrm{gr}\ge1$ (see Fig.\,\ref{fig2}) and $c^2\le1$,
solitons of arbitrary height travel slower than any small
perturbations of the flat-interface state. The maximal speed of
solitons, $c_{\max}=1$, coincides with the minimal group velocity
of linear waves. This yields notable information on the system
dynamics. Fast solitons with $c$ tending to $1$ from below are
extended and have a small height (see equation~(\ref{eq20})), while
envelopes of long linear waves propagate with velocity
$v_\mathrm{gr}$ tending to $1$ from above. This means that
envelopes of small-height soliton packages travel faster than
solitons in these packages. The issue of the generality
of situations where the ranges of the possible soliton velocities
and the group velocities of linear waves do not overlap but only
touch each other is interestingly addressed
in~\cite{Akylas-1993,Longuet-Higgins-1993} from the view point of
emission of wave packages by the soliton (or the impossibility of such
an emission).

\subsection{Stability of solitons}
The stability properties of solitons in the `plus' Boussinesq
equation were addressed in
literature~\cite{Manoranjan-etal-1988,Bona-Sachs-1988,Liu-1993};
in~\cite{Bona-Sachs-1988} the solitons with $1/2\le c\le1$ were
proved to be stable and in~\cite{Liu-1993} the solitons with
$c<1/2$ were proved to be unstable. One can add more subtle
details to this information: the spectrum of Lyapunov exponents
(exponential growth rate) and the dependence of the scenario of
nonlinear growth of perturbations on the initial perturbation.

The problem of linear stability of the soliton $\eta_0(x-ct)$ to
perturbations $\big(e^{\lambda t}\eta_1(x_1),e^{\lambda
t}\varphi_1(x_1)\big)$ in the copropagating reference frame
$x_1=x-ct$ reads
\begin{eqnarray}
\lambda\varphi_1+c\varphi_1^{\prime}&=&
 \eta_1^{\prime\prime}+3\eta_0(x_1)\,\eta_1-\eta_1\,,
\label{eq22}
\\
\lambda\eta_1+c\eta_1^{\prime}&=&-\varphi_1^{\prime\prime}
\label{eq23}
\end{eqnarray}
with boundary conditions
\begin{equation}
\eta_1(\pm\infty)=\varphi_1(\pm\infty)=0\,.
\label{eq24}
\end{equation}

\begin{figure}[t]
\center{
{\sf (a)}\hspace{-7mm}
 \includegraphics[width=0.425\textwidth]%
 {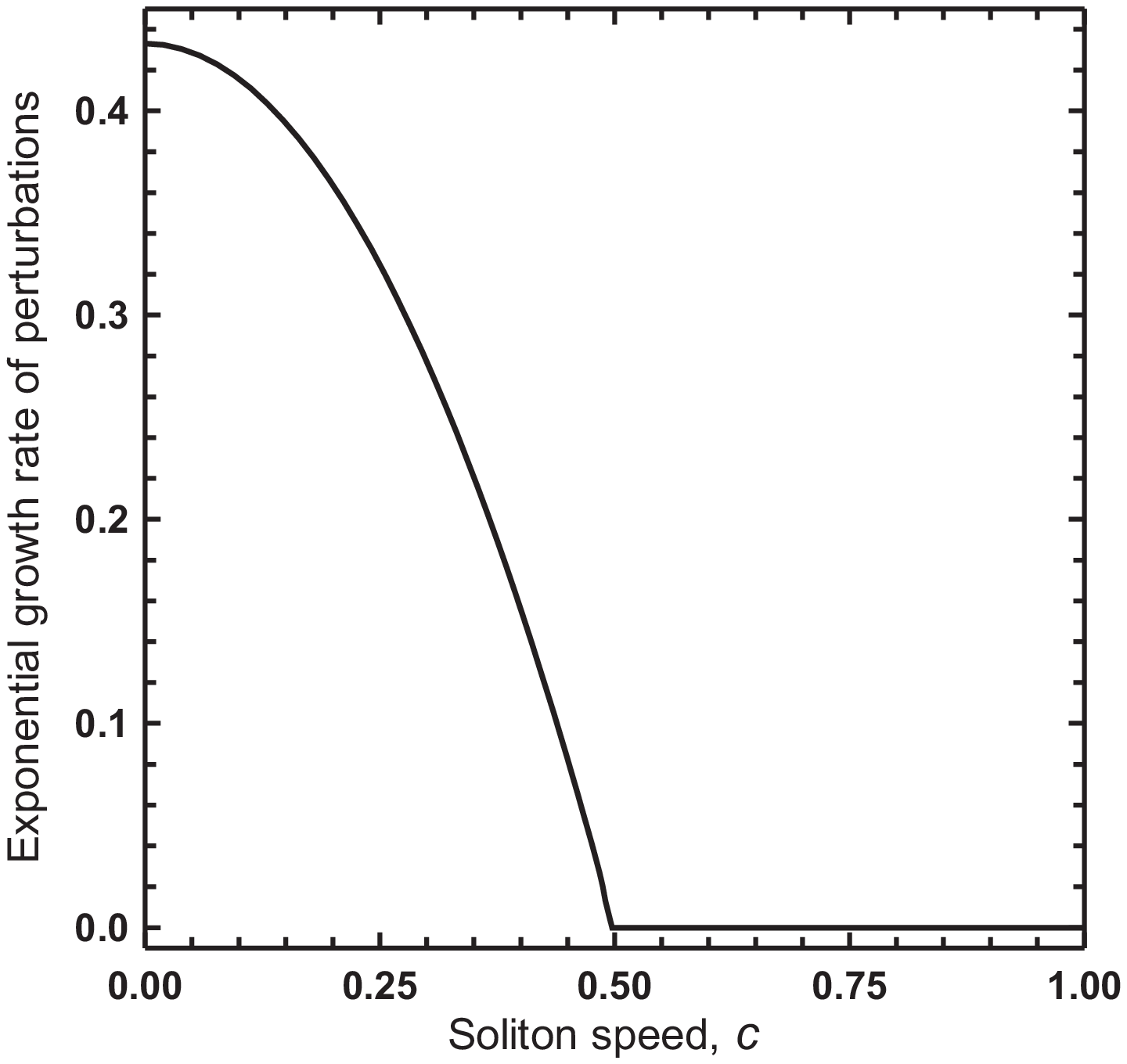}
 \\[25pt]
{\sf (b)}\hspace{-7mm}
 \includegraphics[width=0.425\textwidth]%
 {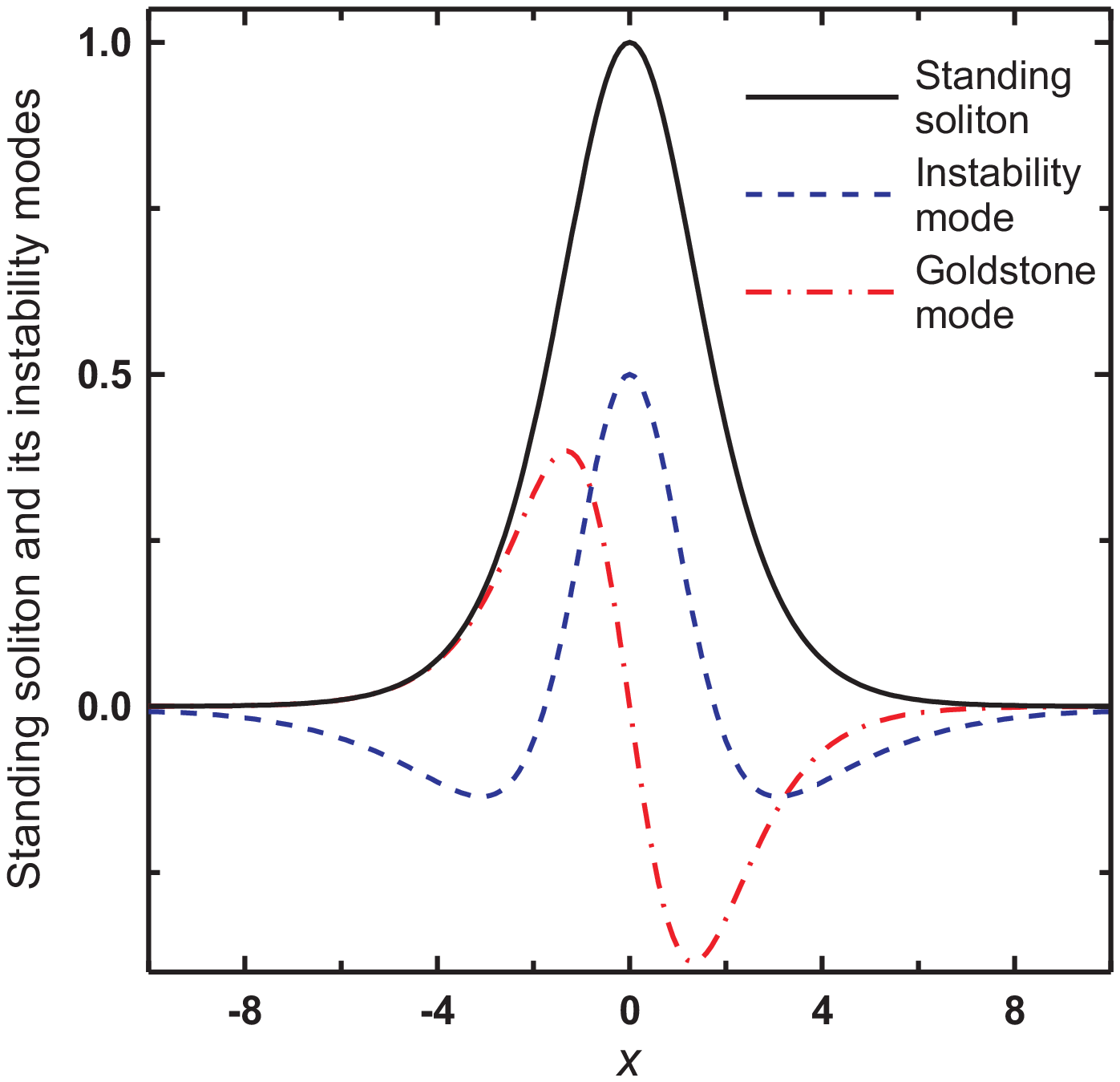}}

  \caption{
(a):~Exponential growth rate $\Re(\lambda)$ of perturbations of
soliton as a function of the soliton speed $c$. (b):~The
instability mode and the Goldstone mode of neutral stability
(invariance to the shifts of soliton in space) of the standing
soliton ($c=0$).
 }
  \label{fig4}
\end{figure}

The eigenvalue problem~(\ref{eq22})--(\ref{eq24}) was solved
numerically with employment of the shooting method. The spectra of
eigenvalues $\lambda$ for different $c$ are plotted in Fig.\,\ref{fig3}
and the first two eigenmodes of perturbations of the standing
soliton ($c=0$) are plotted in Fig.\,\ref{fig4}(b). In Fig.\,\ref{fig4}(a),
one can see the exponential growth rate $\Re(\lambda)$ of
perturbations; the standing and slow solitons with $c<0.5$ are
unstable, while the fast solitons with $c\ge 0.5$ are stable.

\begin{figure}[t]
\center{
{\sf (a): $c=0.3$}\hspace{-15mm}
 \includegraphics[width=0.45\textwidth]%
 {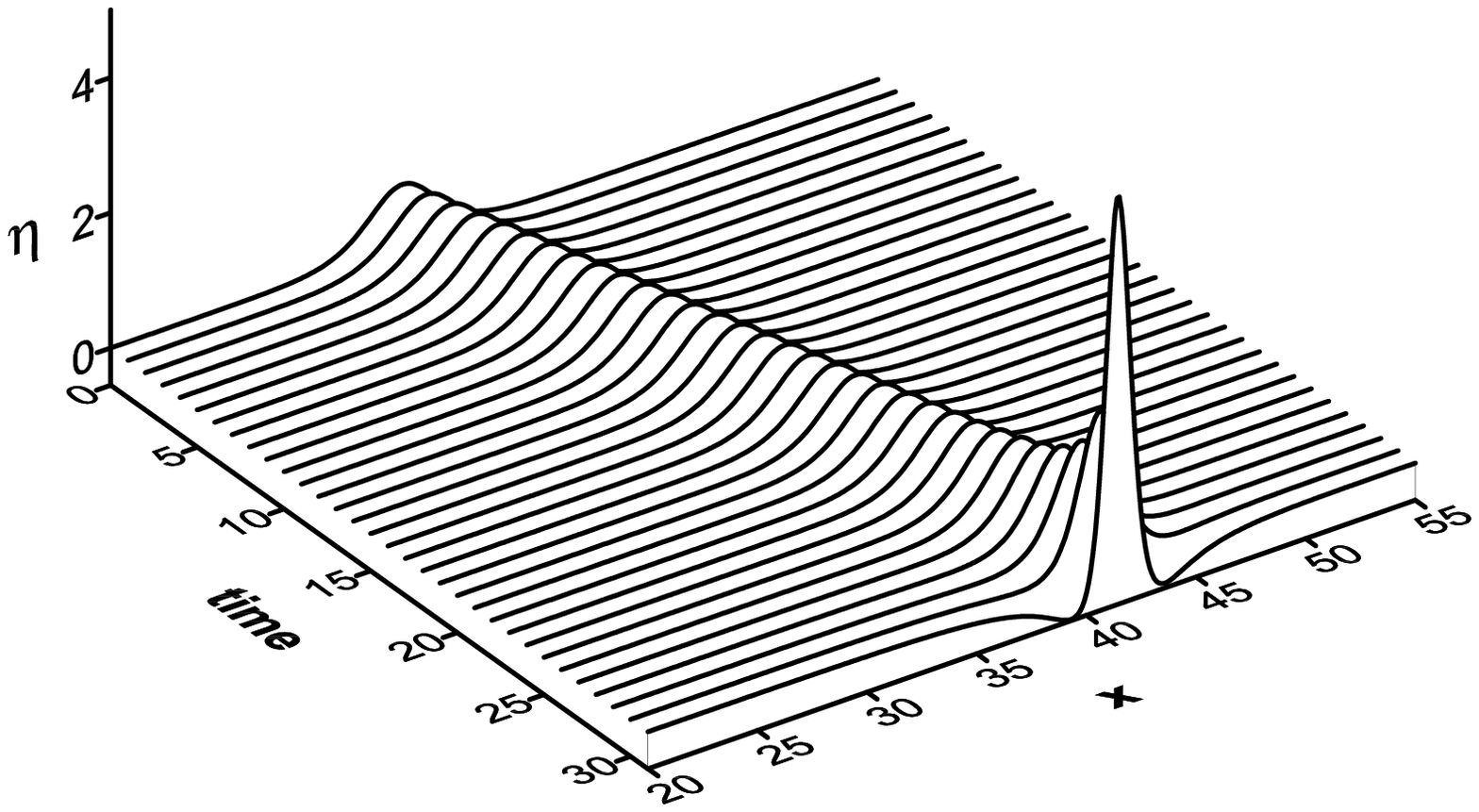}
 \\[25pt]
{\sf (b): $c=0.3$}\hspace{-15mm}
 \includegraphics[width=0.45\textwidth]%
 {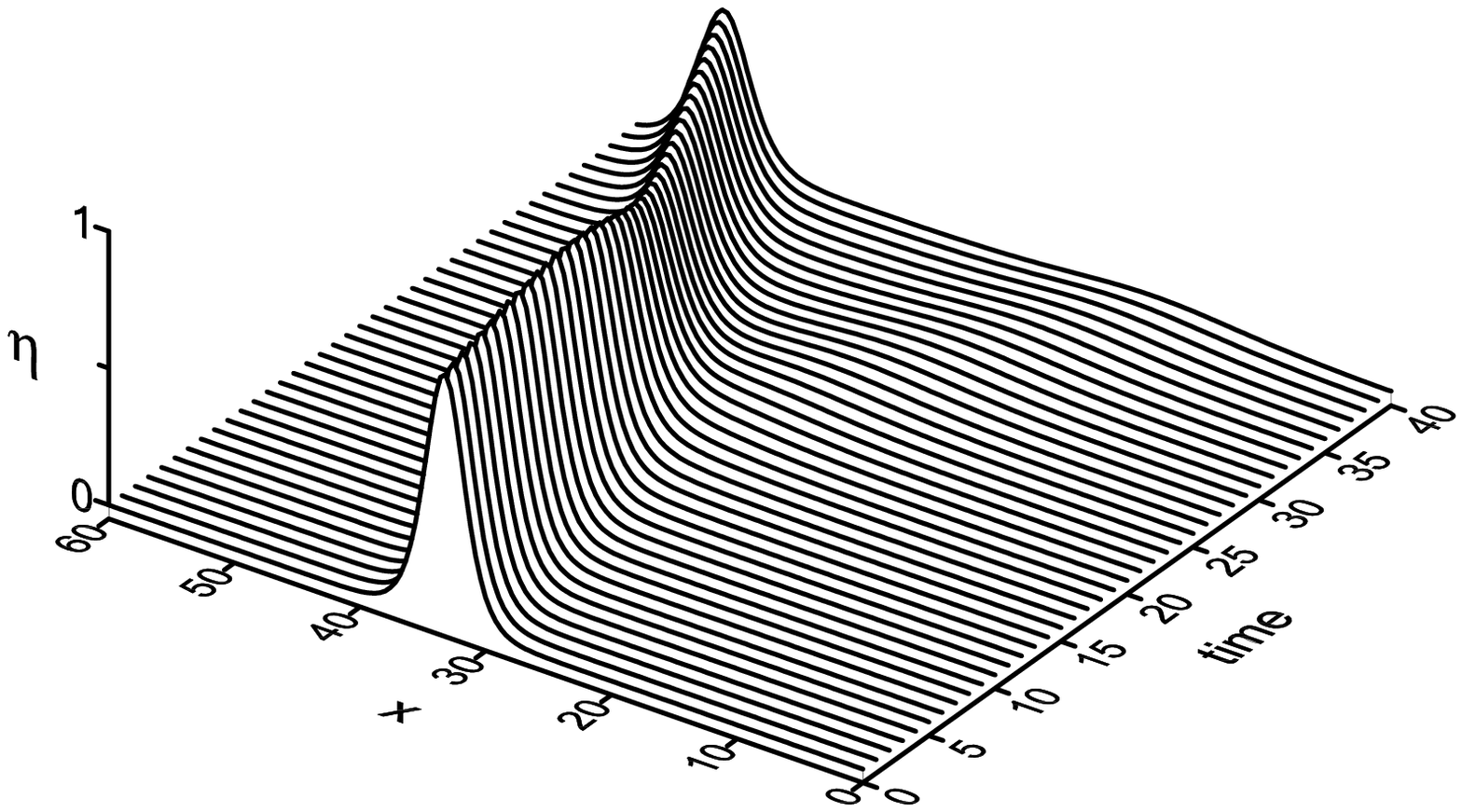}
 \\[25pt]
{\sf (c): $c=0.1$}\hspace{-15mm}
 \includegraphics[width=0.45\textwidth]%
 {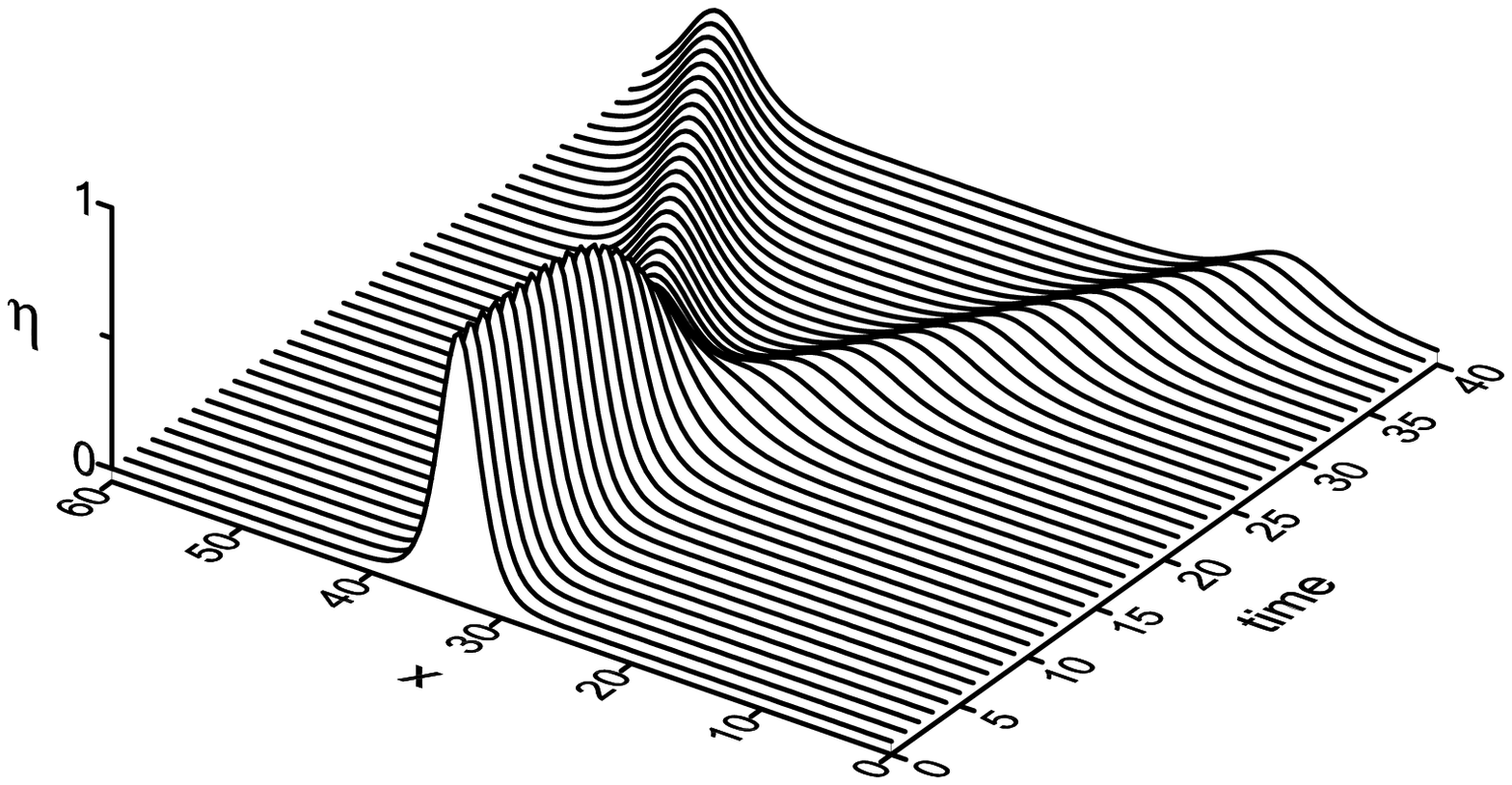} }

  \caption{
(a):~The unstable soliton (\ref{eq20}) of $c=0.3$ with tiny
positive perturbation explodes leading to the formation of a
finite amplitude relief, possibly layer rupture. (b):~The same
unstable soliton ($c=0.3$) with tiny negative perturbation falls
apart into two fast stable solitons. (c):~Falling-apart of the
unstable soliton with $c=0.1$.
 }
  \label{fig5}
\end{figure}

The scenarios of evolution of unstable solitons were observed
numerically by means of direct numerical simulation of the dynamic
system~(\ref{eq14})--(\ref{eq15}) with the finite difference
method in an $x$-domain of length $200$ with periodic boundary
conditions and the space step size $h_x=0.05$.~\footnote{The fact,
that unperturbed unstable analytical solution used as initial
conditions could persist for up to $100$ time units, suggests the
discretisation and numerical integration schemes introduce quite a
small inaccuracy into the system simulation.} As
in~\cite{Manoranjan-etal-1984,Manoranjan-etal-1988,Bogdanov-Zakharov-2002},
two possible scenarios were observed: (i)~soliton explosion with
formation of a finite amplitude relief or, possibly, layer
rupture; (ii)~falling-apart of the soliton into two stable
solitons (Fig.\,\ref{fig5}). Since the phase space of the system is
infinite-dimensional, the problem of discrimination of the
initial perturbations leading to explosion and those leading to
falling-apart may be generally nontrivial. However, in Fig.\,\ref{fig3}
one can see that there is only one instability mode for $c<1/2$
and the nonlinear evolution of perturbations turns out to depend
only on projection of the small initial perturbation on this
unstable direction. If the instability mode is normalized in such
a way that $\eta_1(x_1=0)>0$ (cf.\ Fig.\,\ref{fig4}(b)), the initial
perturbations with a positive scalar product with the instability
mode lead to explosion, while the perturbations with a negative
scalar product lead to falling-apart into two stable solitons.

\begin{figure*}[t]
\center{
 \includegraphics[width=0.68\textwidth]%
 {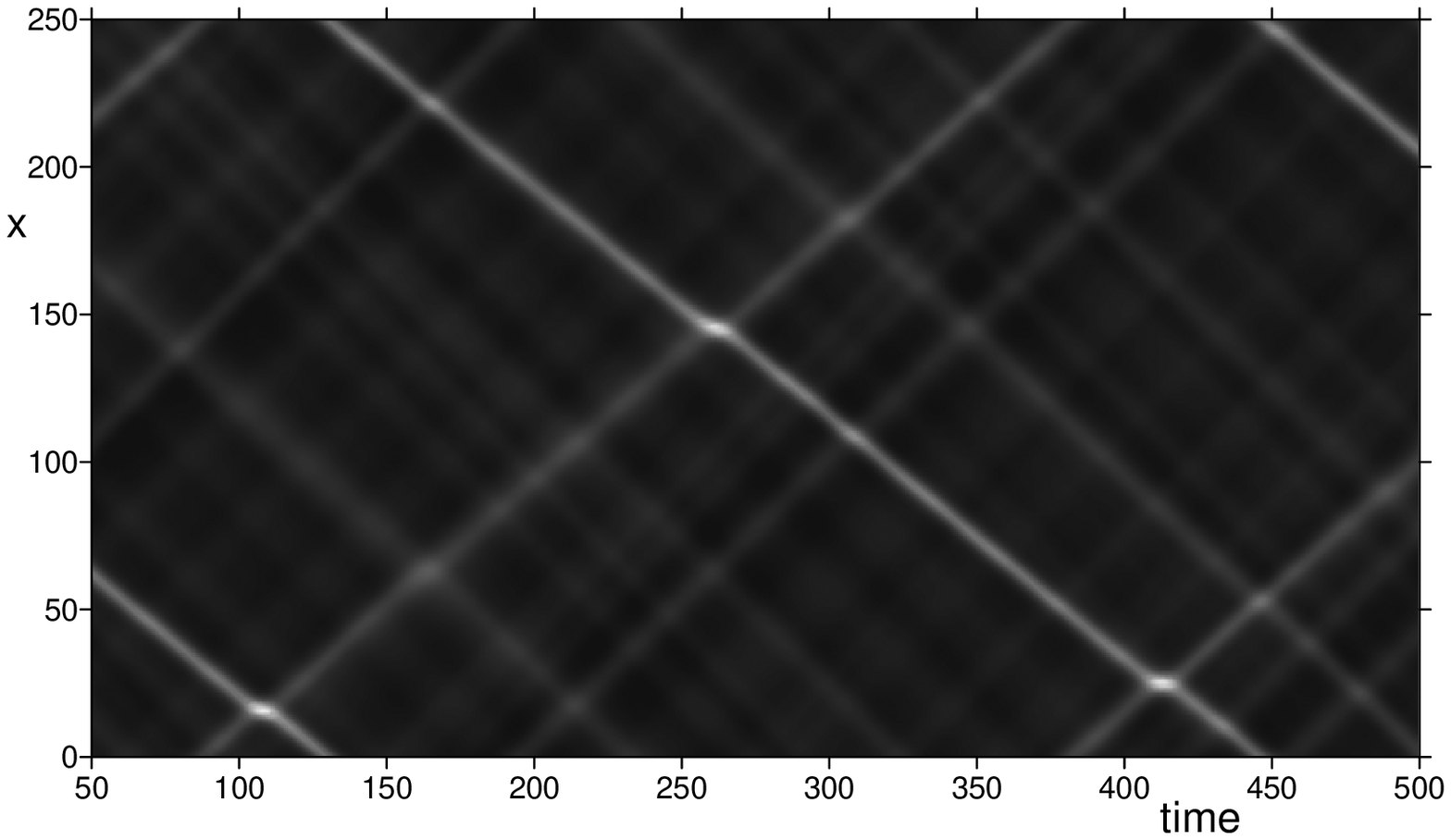}}

  \caption{
Sample evolution of the dynamic system~(\ref{eq14})--(\ref{eq15})
with arbitrary smooth initial conditions. The dynamics can be
viewed as a kinetics of a gas of stable (fast) solitons.
 }
  \label{fig6}
\end{figure*}

\subsection{Soliton gas}
In Fig.~\ref{fig6}, a sample of the system dynamics from arbitrary
initial conditions is presented in domain $x\in[0;250]$ with
periodic boundary conditions. One can see that, beyond the
locations of formation of
singularities~\cite{Goldobin-etal-EPL-2014-solitons}, this
dynamics can be well treated as the kinetics of a gas of stable
solitons. For wave dynamics in soliton-bearing systems,
statistical physics approaches which describe the dynamics of
dense soliton gases can be developed~\cite{El-Kamchatnov-2005}.

Knowing that the system dynamics can be viewed as a kinetics of a
soliton gas, we can readdress the question of relationships
between the group velocity of linear waves and the speed of
solitons. The fast solitons with $c\to1$ have height
$\eta_\mathrm{max}=[1-c^2]\to0$ and width
$\delta\propto1/\sqrt{1-c^2}\to\infty$, i.e.\ must obey the laws
established for the linear waves with wavenumber $k\to0$.
Meanwhile for the latter waves we know the group velocity
$v_\mathrm{gr}=1+(3/2)k^2+\mathcal{O}(k^4)$ [see
Eq.\,(\ref{eq18})]. Thus, the envelopes of traveling solitons
always travel faster than these solitons. For an envelope of a
nearly monochromatic wave the possibility of such a behavior is
obvious, while for a gas of quasi-particles additional
explanations are needed. For waves of density of
quasi-particles (correspond to waves of envelope) it is actually
possible to travel faster than the fastest particles if these
particles have a finite ``collision diameter''. (For a better
intuition on this, one can imagine the elastic collision of two
hard spheres moving along a line. These spheres exchange their
momentums from the distance equal to the sum of their radii. If
they are identified only by the momentum, they effectively jump
for the distance of their interaction and proceed to move with the
initial momentums.) Indeed, colliding copropagating solitons
exchange their momentums, which means they efficiently exchange
their locations, not crossing one another, but approaching for a
certain finite distance, the collision diameter. Thus the momentum
efficiently jumps in the direction of soliton motion for this
distance and, during one and the same time interval, the wave in a
gas can cover a longer distance than the gas quasi-particles.

As soon as one can speak of the soliton density waves in a soliton
gas, the question of the criterion for this gas to be considered
as a continuous medium or a vacuum arises. For instance, it is
obvious that one cannot speak of density waves or envelope
waves for a system state with a single soliton; this is a
vacuum. Whereas, for a continuous medium the concept of the group
velocity should work well. Let us consider a gas of solitons with
characteristic width $\delta\gg1$, i.e., quasi-particle speed
$c^2=1-\delta^{-2}$ [see Eq.\,(\ref{eq20})]. The signal transfer
speed due to collisions (as described in the paragraph above) is
larger that the particle speed, according to $v_\mathrm{gr}\approx
c/(1-n\delta)$, where $n$ is the soliton number density.
Mathematically, the criterion for $n$ is of interest. For us to
be able to consider the soliton gas as a continuous medium, with
the group velocity featured by Eq.\,(\ref{eq18}), $v_\mathrm{gr}$
should at least reach $1$. Hence, $c/(1-n_\mathrm{min}\delta)=1$,
which can be rewritten as
$1-\delta^{-2}/2=1-n_\mathrm{min}\delta$, and one finally finds
\begin{equation}
n_\mathrm{min}\sim\delta^{-3}\,.
\label{eq:g01}
\end{equation}
Interestingly, the last equation means that the maximal
characteristic inter-soliton distance
$\delta_\ast=1/n_\mathrm{min}$ for the gas to be a continuous
medium but not a vacuum scales with the soliton width $\delta$ as
\begin{equation}
\delta_\ast\sim\delta^3\,.
\label{eq:g02}
\end{equation}

\section{Conclusion}
\label{sec_concl}
We have considered the dynamics of patterns on the internal
surface of the horizontal two-layer system of inviscid fluids
subject to tangential vibrations. For thin layers
($h<\sqrt{3\alpha/[(\rho_2-\rho_1)g]}$) the
instability is known to be long-wavelength and
subcritical~\cite{Lyubimov-Cherepanov-1987,Zamaraev-Lyubimov-Cherepanov-1989}.
The governing equations for long-wavelength patterns below the linear
instability threshold have been derived---equation system
(\ref{eq14})--(\ref{eq15})---allowing for the first time
theoretical analysis for time-dependent patterns in the system and
for stability of time-independent (quasi-steady) patterns.
We note that the stability analysis for the only time-independent
localized patterns in the system, standing solitons, has revealed
them to be unstable.

The system dynamics is found to be governed by dynamic
system~(\ref{eq14})--(\ref{eq15}) which is equivalent to the
`plus' Boussinesq equation. For dynamic
system~(\ref{eq14})--(\ref{eq15}), one-parametric family of
localized solutions of time-independent profile, solitons, exists
(equation~(\ref{eq20})). These solitons are up-standing
embossments of the interface (cf.\ black curve in
Fig.\,\ref{fig4}(b)) and are parameterized by the soliton speed $c$
only, which varies from $c=0$ (the tallest and sharpest solitons)
to $c=1$ (solitons with width tending to infinity and height
tending to zero). The standing and slow solitons ($c<1/2$) are
unstable~\cite{Liu-1993}, while the fast solitons ($c\ge 1/2$) are
stable~\cite{Bona-Sachs-1988}. The group velocity of linear waves
in the system is $v_\mathrm{gr}\ge1$, meaning that all the solitons
travel more slowly than any wave packages of small perturbations of the
flat-interface state.

Two scenarios of development of the instability of slow solitons
are possible, depending on the initial perturbation: explosion
(probably leading to further layer rupture) and splitting into
a pair of fast stable solitons. No other localized waves have been
detected with direct numerical simulation, meaning that this
one-parameter family of solitons is the only localized waves in
the system. The system dynamics can be fully represented as
the kinetics of gas of solitons before an explosion (and after it).

It is not possible to compare our results to the results presented
by Wolf~\cite{Wolf-1961,Wolf-1970} in detail. Wolf presented the
wave patterns of the interface for the inverted state (the heavy
liquid above the light one) above the linear instability threshold
of the flat-interface non-inverted state. Meanwhile, we consider
waves on the interface for the non-inverted state below the
threshold, and our non-trivial findings pertain specifically to
this case but not to the case of the inverted stratification.

We are thankful to Dr.\ Maxim V.\ Pavlov and Dr.\ Takayuki
Tsuchida for their useful comments on the work and drawing our
attention to the fact that system (\ref{eq14})--(\ref{eq15}) is
identical to the `plus' Boussinesq equation. We thank Prof.\
Jeremy Levesley for his help with manuscript preparation. The work
has been supported by the Russian Science Foundation (grant
no.~14-21-00090).

To the memory of our teachers and friends A.~A.\ Cherepanov, D.~V.\ Lyubimov, and S.~V.\ Shklyaev.


\begin{thebibliography}{25}

\bibitem{Wolf-1961}
G.\ H.\ Wolf,
 {\em The dynamic stabilization of the Rayleigh-Taylor instability and the corresponding dynamic equilibrium},
\newblock Z.\ Phys. {\bf 227}, 291 (1969).

\bibitem{Wolf-1970}
G.\ H.\ Wolf,
 {\em Dynamic Stabilization of the Interchange Instability of a Liquid-Gas Interface},
\newblock Phys.\ Rev.\ Lett. {\bf 24}, 444 (1970).

\bibitem{Lyubimov-Cherepanov-1987}
D.\ V.\ Lyubimov and A.\ A.\ Cherepanov,
 {\em Development of a steady relief at the interface of fluids in a vibrational field},
\newblock Fluid Dynamics {\bf 21}, 849 (1986).

\bibitem{Khenner-Lyubimov-Shotz-1998}
D.\ V.\ Lyubimov, M.\ V.\ Khenner, and M.\ M.\ Shotz,
 {\em Stability of a Fluid Interface Under Tangential Vibrations},
\newblock Fluid Dynamics {\bf 33}, 318 (1998).

\bibitem{Khenner-etal-1999}
M.\ V.\ Khenner, D.\ V.\ Lyubimov, T.\ S.\ Belozerova, and B.\ Roux,
 {\em Stability of Plane-Parallel Vibrational Flow in a Two-Layer System},
\newblock European Journal of Mechanics B/fluids {\bf 18}, 1085 (1999).

\bibitem{Zamaraev-Lyubimov-Cherepanov-1989}
A.\ V.\ Zamaraev, D.\ V.\ Lyubimov, and A.\ A.\ Cherepanov,
 {\em On equlibrium shapes of the interface between two fluids in vibrational field},
\newblock in {\em Hydrodynamics and Processes of Heat and Mass Transfer}
 (Ural Branch of Acad.\ of Science of USSR, Sverdlovsk, 1989), pp.\ 23--26.
 [In Russian. Since the translation of this paper into English in not available in the literature, it may be suitable to notice that the results of this work related to the problem we consider can be as well deduced from~\cite{Lyubimov-Cherepanov-1987}.]


\bibitem{Shklyaev-Alabuzhev-Khenner-2009}
S.\ V.\ Shklyaev, A.\ A.\ Alabuzhev, and M.\ Khenner,
 {\em Influence of a longitudinal and tilted vibration on stability and dewetting of a liquid film},
\newblock Phys.\ Rev.\ E {\bf 79}, 051603 (2009).

\bibitem{Benilov-Chugunova-2010}
E.\ S.\ Benilov and M.\ Chugunova,
 {\em Waves in liquid films on vibrating substrates},
\newblock Phys.\ Rev.\ E {\bf 81}, 036302 (2010).

\bibitem{Thiele-Vega-Knobloch-2006}
U.\ Thiele, J.\ M.\ Vega, and E.\ Knobloch,
 {\em Long-wave Marangoni instability with vibration},
\newblock J.\ Fluid Mech. {\bf 546}, 61 (2006).

\bibitem{Nepomnyashchy-Simanovskii-2013}
A.\ A.\ Nepomnyashchy and I.\ B.\ Simanovskii,
 {\em The influence of vibration on marangoni waves in two-layer films},
\newblock J.\ Fluid Mech. {\bf 726}, 476 (2013).

\bibitem{Goldobin-Lyubimov-2007}
D.\ S.\ Goldobin and D.\ V.\ Lyubimov,
 {\em Soret-Driven Convection of Binary Mixture in a Horizontal Porous Layer in the Presence of a Heat or Concentration Source},
\newblock JETP {\bf 104}, 830 (2007).

\bibitem{Watson-1964}
E.\ J.\ Watson,
 {\em The radial spread of a liquid jet over a horizontal plane},
\newblock J.\ Fluid Mech. {\bf 20}, 481 (1964).

\bibitem{Boussinesq-1872}
J.\ Boussinesq,
 {\em Th{\'e}orie des ondes et des remous qui se propagent le long d'un canal rectangulaire horizontal, en communiquant au liquide contenu dans ce canal des vitesses sensiblement pareilles de la surface au fond},
\newblock Journal de Math{\'e}matiques Pures et Appliqu{\'e}es {\bf 17}, 55 (1872).

\bibitem{Choi-Camassa-1999}
W.\ Choi and R.\ Camassa,
 {\em Fully nonlinear internal waves in a two-fluid system},
\newblock J.\ Fluid Mech. {\bf 396}, 1 (1999).

\bibitem{Manoranjan-etal-1984}
V.\ S.\ Manoranjan, A.\ R.\ Mitchell, and J.\ L.\ Morris,
 {\em Numerical solutions of the good Boussinesq equation},
\newblock SIAM J.\ Sci.\ Statist.\ Comput. {\bf 5}(4), 946 (1984).

\bibitem{Manoranjan-etal-1988}
V.\ S.\ Manoranjan, T.\ Ortega, and J.\ M.\ Sanz-Serna,
 {\em Soliton and antisoliton interactions in the ``good'' Boussinesq equation},
\newblock J.\ Math.\ Phys. {\bf 29}(9), 1964 (1988).

\bibitem{Bogdanov-Zakharov-2002}
L.\ V.\ Bogdanov and V.\ E.\ Zakharov,
 {\em The Boussinesq equation revisited},
\newblock Physica D {\bf 165}, 137 (2002).

\bibitem{Bona-Sachs-1988}
J.\ L.\ Bona and R.\ L.\ Sachs,
 {\em Global existence of smooth solutions and stability of solitary waves for a generalized Boussinesq equation},
\newblock Comm.\ Math.\ Phys. {\bf 118}, 15 (1988).

\bibitem{Liu-1993}
Y.\ Liu,
 {\em Instability of solitary waves for generalized Boussinesq equations},
\newblock J.\ Dynam.\ Differential Equations {\bf 5}, 537 (1993).

\bibitem{Akylas-1993}
T.\ R.\ Akylas,
 {\em Envelope solitons with stationary crests},
\newblock Phys.\ Fluids {\bf 5}, 789 (1993).

\bibitem{Longuet-Higgins-1993}
M.\ S.\ Longuet-Higgins,
 {\em Capillary--gravity waves of solitary type and envelope solitons on deep water},
\newblock J.\ Fluid Mech. {\bf 252}, 703 (1993).

\bibitem{Goldobin-etal-EPL-2014-solitons}
D.\ S.\ Goldobin, K.\ V.\ Kovalevskaya, and D.\ V.\ Lyubimov,
 {\em Elastic and inelastic collisions of interfacial solitons and integrability of a two-layer fluid system subject to horizontal vibrations},
\newblock EPL (Europhys.\ Lett.) {\bf 108}, 54001 (2014).

\bibitem{El-Kamchatnov-2005}
G.\ El and A.\ Kamchatnov,
 {\em Kinetic Equation for a Dense Soliton Gas},
\newblock Phys.\ Rev.\ Lett. {\bf 95}, 204101 (2005).

\end{thebibliography}
\end{document}